\documentclass[prl,twocolumn,amsmath,amssymb,superscriptaddress,showpacs]{revtex4}
\usepackage{graphics}
\usepackage{epsfig}
\usepackage{bbm}
\usepackage[latin1]{inputenc} 

\newcommand{\be}{\begin{equation}}
\newcommand{\ee}{\end{equation}}
\newcommand{\eea}{\end{eqnarray}}
\newcommand{\bea}{\begin{eqnarray}}

\newcommand{\mean}[1]{\ensuremath{\langle{#1}\rangle}}

\newcommand{\eins}{\ensuremath{\mathbbm 1}}
\newcommand{\II}{\ensuremath{\mathbbm I}}
\newcommand{\qed}{\ensuremath{\hfill \Box}}
\newcommand{\fa}{\ensuremath{\mathfrak{a}}}
\newcommand{\fb}{\ensuremath{\mathfrak{b}}}

\newcommand{\WW}{\ensuremath{\mathcal{W}}}
\newcommand{\FF}{\ensuremath{\mathcal{F}}}
\newcommand{\GG}{\ensuremath{\mathcal{G}}}
\newcommand{\NL}{\ensuremath{\mathcal{X}}}

\newcommand{\HH}{\ensuremath{\mathcal{H}}}
\newcommand{\BB}{\ensuremath{\mathcal{B}}}

\newcommand{\ketbra}[1]{\ensuremath{| #1 \rangle \langle #1 |}}

\newcommand{\ket}[1]{\ensuremath{|#1\rangle}}

\newcommand{\bra}[1]{\ensuremath{\langle#1|}}

\newcommand{\braket}[2]{\ensuremath{\langle #1|#2\rangle}}

\newcommand{\kommentar}[1]{}
\renewcommand{\vr}{\ensuremath{\varrho}}
\renewcommand{\eins}{\ensuremath{\openone}}

\begin{document}
\title{Nonlinear entanglement witnesses}
\date{\today}
\begin{abstract}
Entanglement detection typically relies on linear inequalities 
for mean values of certain observables (entanglement witnesses), 
where violation indicates entanglement. We provide a general method 
to improve any of these 
inequalities  for bipartite systems via nonlinear expressions. The 
nonlinearities are of different orders and can be directly measured 
in experiments, often without any extra effort. 
\end{abstract}

\author{Otfried G\"uhne}
\affiliation{Institut f\"ur Quantenoptik und Quanteninformation, 
\"Osterreichische Akademie der Wissenschaften, 6020 Innsbruck, Austria}

\author{Norbert L\"utkenhaus}
\affiliation{Quantum Information Theory Group, 
Institut f\"ur Theoretische Physik I, 
and Max-Planck Research Group, Institute of Optics, Information and Photonics, 
Universit\"{a}t Erlangen-N\"{u}rnberg, Staudtstra{\ss}e 7/B2, 91058 Erlangen, Germany}
\affiliation{Institute for Quantum Computing, University of Waterloo, 
200 University Avenue West, Waterloo, Ontario N2L 3G1, Canada}	

\maketitle

Entanglement detection is one of the fundamental problems
in quantum information science. On the one hand, it is 
crucial for experiments, since the question whether a 
produced state is entangled or not may decide whether a 
given experiment has been successful or not. On the other
hand, it is also a challenging task for theoreticians, since
the separability problem is, despite a lot of progress in the 
last years, essentially not solved. 

In most of the experiments, so called entanglement witnesses 
are used for entanglement verification 
\cite{witnessexp, mohamed, witnessth, ppthoro, optimization, sanpera, pmap}. 
These are linear inequalities for mean values of 
certain observables, if the inequality is violated, 
the state must be entangled. For instance, Bell 
inequalities are such linear inequalities and may 
hence be viewed as entanglement witnesses. 
The question whether one can use {\it nonlinear} 
inequalities for entanglement detection has also been 
discussed and many nonlinear criteria are known 
\cite{nonlinear, lurs, uncertaintyprl, hyllus, aip, uffink}. 

In this situation it is natural to ask whether it is 
possible to improve a {given} linear witness by some 
nonlinear expression. 
For discrete systems two examples for such an improvement 
are known. Uffink showed that certain Bell inequalities 
can be improved by nonlinear expressions \cite{uffink}. 
Recently, Hofmann and Takeuchi proposed separability 
conditions called ``local uncertainty relations'' 
\cite{lurs}, which can improve witnesses in some cases 
\cite{aip}. For continuous variable systems it is known
how to summarize certain families of linear inequalities 
to a single nonlinear one \cite{hyllus}.

In this Letter we provide a method which allows to improve 
{\it all} entanglement witnesses for discrete bipartite 
systems by nonlinear expressions. Surprisingly, this works 
also for optimal entanglement witnesses, that is, witnesses 
where no other linear witness is stronger. Due to the 
simplicity of our method the nonlinear witnesses may 
be used from the beginning, without considering linear 
witnesses anymore. Our method allows to calculate such 
nonlinear witnesses to an arbitrary order. 
We discuss in detail the strength of our constructions for 
the case of two qubits. Interestingly, when implemented in 
an experiment, the nonlinear expressions in our constructions 
often require measurement of the same observables as the 
original witness. This allows for an improved entanglement 
detection without any extra effort.

Without loosing generality, we consider witnesses $\WW$ 
as observables with a positive mean value on all separable 
states, thus a negative expectation value signals the presence 
of entanglement. Our aim is to find a nonlinear witness $\FF$ 
as a functional of the type $\FF(\vr) = Tr(\WW \vr) - \NL(\vr)$
which still should be positive on all separable states. 
Typically, the nonlinearity $\NL(\vr)$ 
will be a quadratic polynomial of certain expectation 
values. We consider  
only $\FF(\vr)$ which are {\it strictly stronger} 
than the witness $\WW$. That is, we require that   
$\FF(\vr)$ detects all the states that are detected 
by $\WW$ and some states in addition. 

Let us explain the main idea for the case of witnesses coming 
from the separability criterion of the positivity of the partial 
transpose (PPT) \cite{ppt}. By definition, a quantum state $\vr$  
shared between Alice and Bob is separable if it can be written as a 
mixture of product states, that is 
\be
\vr= \sum_{i} p_i \vr^A_i \otimes \vr^B_i, 
\label{sepdef}
\ee
where $p_i \geq 0$ and $\sum_i p_i =1.$ Then, given
a quantum state 
$
\vr=
\sum_{ij,kl} \vr_{ij,kl} 
\ket{i}\bra{j}\otimes\ket{k}\bra{l}
$
its partial transpose with respect to Bob's system is defined 
as
$
\vr^{T_B}=
\sum_{ij,kl} \vr_{ij,lk} \ket{i}\bra{j}\otimes\ket{k}\bra{l}.
$
If $\vr$ is a separable state, it can be easily seen that 
the partial transpose is positive, $\vr^{T_B}\geq 0.$ Thus, 
if for a state the partial transpose is not positive 
($\vr$ is NPT), the state must be entangled. Indeed, 
it has been shown \cite{ppthoro} that for $2 \times 2$ 
and $2 \times 3$ systems a state is PPT if and only if 
it is separable, while for other dimensions there are 
also PPT entangled states. 

For any NPT state $\vr_0$ we find that $\vr_0^{T_B}$ has a negative 
eigenvalue $\lambda_-$ and a corresponding eigenvector 
$\ket{\phi}.$  An entanglement witness for this state 
is then
\be
\WW = \ketbra{\phi}^{T_B}.
\label{wit1}
\ee
Indeed, due to the identity $Tr(XY^{T_B})=Tr(X^{T_B}Y)$ for 
arbitrary matrices $X,Y,$ we have $Tr(\vr_0\WW)=\lambda_- < 0$ 
while for separable (and hence PPT) states 
$Tr(\vr\WW)=\bra{\phi}\vr^{T_B}\ket{\phi}\geq 0$ holds. Note 
that the witness in Eq.~(\ref{wit1}) is by no means specific: 
since the PPT criterion is necessary and sufficient for low 
dimensions, witnesses of this type suffice to detect all states 
in these systems.  Furthermore, such witnesses can be shown to 
be optimal \cite{optimization}.

To improve the witness from Eq.~(\ref{wit1}) with nonlinear 
terms, first note that a functional like $\mean{X}\mean{X^\dagger}$
is convex (see Lemma 1 in Ref.~\cite{uncertaintyprl}). This implies 
that a functional like 
$\GG = \mean{A} - \sum_i \alpha_i \mean{X_i}\mean{X_i^\dagger}$
is concave in the state. That is, if $\vr=\sum_k p_k \vr_k$ 
is a convex combination of some states, then 
$\GG(\vr) \geq \sum_k p_k \GG(\vr_k).$ 

Let us assume that we have chosen 
$ A= \ketbra{\phi}, X_i = \ket{\phi}\bra{\psi_i}$
for an arbitrary $\ket{\psi_i}$ and take a separable 
state $\vr.$ Then $\vr^{T_B}$ 
is again separable and can 
be written as a convex combination of product 
states,  $\vr^{T_B}=\sum_k p_k \ketbra{a_k b_k}.$ 
For $\ket{\chi}=\ket{a_k b_k}$ we have
\be
\GG(\ketbra{\chi}) = 
\braket{\chi}{\phi}\braket{\phi}{\chi}
\cdot
\underbrace{
\big[
1 - \sum_i \alpha_i \braket{\chi}{\psi_i}\braket{\psi_i}{\chi}
\big].}_{=: P(\chi)}
\label{nl1} 
\ee
Thus, if the polynomial $P(\chi)$ is positive on all
product states, by concavity the functional $\GG$ is 
positive on all separable $\vr^{T_B}$. Then, with the 
chosen $X_i,$
\be
\FF = 
\mean{\ketbra{\phi}^{T_B}} 
- \sum_i \alpha_i \mean{X_i^{T_B}}\mean{(X_i^{T_B})^\dagger}
\label{nl2}
\ee 
is  a nonlinear improvement of the witness 
$\WW=\ketbra{\phi}^{T_B}.$ It is important 
to note that a term like
$\mean{X^{T_B}} \mean{(X^{T_B})^\dagger}$
in Eq.~(\ref{nl2}) is easily accessible in experiments,  
even if $X^{T_B}$ is non-Hermitean. Namely, 
we can write $X^{T_B}=H+i \cdot A$ as a sum of 
its Hermitean and anti-Hermitean part, where 
$H$ and $A$ are Hermitean. Then 
$\mean{X^{T_B}} \mean{(X^{T_B})^\dagger}  
= \mean{H}^2+ \mean{A}^2 $ holds.
With this method, we have:

{\bf Observation 1.} (a) Let $\WW = \ketbra{\phi}^{T_B}$ be an 
entanglement witness. We define $X_i=\ket{\phi}\bra{\psi}$
for an arbitrary $\ket{\psi}$ and $s(\psi)$ as the square of 
the largest Schmidt coefficient of $\ket{\psi}.$
Then
\be
\FF^{(1)}(\vr)=
\mean{\ketbra{\phi}^{T_B}} 
- 
\frac{1}{s(\psi)}
\mean{X^{T_B}} \mean{(X^{T_B})^\dagger}
\label{nl3}
\ee
is a nonlinear improvement of $\WW.$
\\
(b) If we  define $X_i=\ket{\phi}\bra{\psi_i}, \;\; i=1,...,K$
with an orthonormal basis $\ket{\psi_i},$ then
\be
\FF^{(2)}(\vr) = \mean{\ketbra{\phi}^{T_B}}
- \sum_{i=1}^K \mean{X_i^{T_B}} \mean{(X_i^{T_B})^\dagger}
\label{nl4}
\ee
is also a nonlinear witness. 

{\it Proof.} 
(a)
The squared
overlap between a state $\ket{\psi}$ and a product 
state is bounded by the maximal squared Schmidt 
coefficient \cite{mohamed}, thus 
$P(\chi)$ in Eq.~(\ref{nl1}) is positive.
(b) For the  $\ket{\psi_i}$ 
we have in Eq.~(\ref{nl1}) 
$\sum_i \braket{\chi}{\psi_i}\braket{\psi_i}{\chi}
= Tr(\ketbra{\chi}) = 1$ which proves the claim. 
$\qed$

This Observation provides a whole class of nonlinear 
improvements, one may pick an arbitrary $\ket{\psi}$ 
and compute the corresponding nonlinearity. In special 
situations one can also adjust the improvement: if an 
experiment produces the state $\vr(p)=p \vr_e + (1-p) \vr_n,$
which is a mixture of an entangled $\vr_e$ and some noise 
$\vr_n,$ a given witness 
may detect the states only for $p>p_{\rm crit},$ i.e.,
$Tr(\WW \vr(p_{\rm crit}))=0.$
Now one can choose the $\ket{\psi}$ in a way that the 
nonlinear terms are large at $\vr(p_{\rm crit}),$
which leads to a significant improvement of the 
witness for the states of interest. Concerning the strength 
of the nonlinear improvements we can state:

{\bf Observation 2.} 
(a) Let $\WW = \ketbra{\phi}^{T_B}$ be a witness. 
A state $\vr$ can be detected by a witness of the type
$\FF^{(1)}$ from Eq.~(\ref{nl3}) if and only if 
\be
\bra{\phi}\vr^{T_B}\ket{\phi} < 
\Big[
Tr_B
\big(
\sqrt{Tr_A(\vr^{T_B}\ketbra{\phi}\vr^{T_B})}
\big)
\Big]^2
.
\label{nl5}
\ee
\\ 
(b) A state can be detected by a witness 
of the type $\FF^{(2)}$  from Eq.~(\ref{nl4}) 
if and only if 
\be
\bra{\phi}\vr^{T_B}\ket{\phi} < \bra{\phi}(\vr^{T_B})^2 \ket{\phi}
\label{nl6}
\ee
holds. In this case, the state is detected by all nonlinear witnesses 
of the type $\FF^{(2)}.$ 
\\
(c) If Eq.~(\ref{nl6}) is fulfilled, then also Eq.~(\ref{nl5}) 
holds, thus the witnesses of the type $\FF^{(1)}$ are stronger.
Furthermore, Eqs.~(\ref{nl5}, \ref{nl6}) are never fulfilled for 
PPT states.

{\it Proof.} The proof is given in the Appendix. $\qed$

Let us add that with the same idea also witnesses with 
higher nonlinearities, e.g. of fourth order, can be 
derived. To do so, note first that functionals like
$
\GG = \mean{A}- \alpha \mean{B}\mean{B^\dagger}
-\beta \mean{C}^2\mean{C^\dagger}^2
$
are also concave in the state. If we choose 
$
A=\ketbra{\phi},
B=\ket{\phi}\bra{\psi_1},
C=\ket{\phi}\bra{\psi_2},
$
we arrive at a similar expression as in Eq.~(\ref{nl1}). 
Now, $P(\chi)$ is a polynomial of fourth order that has 
to be positive on all product states \cite{jarek}.

To give an example which we will investigate later, 
let us assume that $\ket{\psi_1},\ket{\psi_2}$
and $\ket{\phi}$ form an orthonormal set, where the biggest 
squared Schmidt coefficient from $\ket{\psi_1},\ket{\psi_2}$
is $1/2$ and the one from $\ket{\phi}$ is $s_{\phi}\geq 1/2.$
Then,  
\bea
\FF^{(3)} &=& \mean{\ketbra{\phi}^{T_B}} - 
(2-\frac{2}{27 s_{\phi}}) \cdot \mean{B^{T_B}}\mean{(B^{T_B})^\dagger}
\nonumber
\\
&&
- \frac{2}{s_{\phi}}  
\mean{C^{T_B}}^2\mean{(C^{T_B})^\dagger}^2
\label{nl7}
\eea
is a nonlinear witness of fourth order. The positivity of the 
polynomial $P(\chi)$ can directly be checked.

It should, however, be noted that higher orders do not necessarily 
provide witnesses, which are much stronger than the witnesses with 
quadratic nonlinearity. 
The reason is that we are not trying to approximate the curvature 
of the convex set of separable states (in which case higher 
polynomials would be clearly an advantage): in our approach, 
we also require the concave functional to be positive on the 
convex set of separable states which makes the situation more 
complicated.

Let us 
discuss an application to two-qubit systems. A generic optimal 
witness for a two-qubit system is
\bea
\WW(\alpha)&=&\ketbra{\phi(\alpha)}^{T_B},
\\
\ket{\phi(\alpha)} &=&\cos(\alpha) \ket{00} + \sin(\alpha) \ket{11}.
\label{nl8}
\eea
It is useful to express all quantities directly in 
expectation values for observables. We use the abbreviations
$s=\sin(\alpha), c=\cos(\alpha), s_2 = \sin(2\alpha), 
c_2=\cos(2\alpha)$ and $x_1 x_2 = \mean{\sigma_x \otimes \sigma_x}, 
y_1 = \mean{\sigma_y \otimes \eins}$ 
etc.~here.
The witness is then
\be
\mean{\WW}= \frac{1}{4} 
\big(
1 + z_1 z_2 + s_2 (x_1x_2+y_1y_2) + c_2 (z_1 + z_2)
\big).
\label{nl9}
\ee
To construct improvements, we choose the four vectors
$
\ket{\psi_{1/2}} = (\ket{00}\pm\ket{11})/\sqrt{2}, 
\ket{\psi_{3/4}} = (\ket{01}\pm\ket{10})/\sqrt{2},
$
where the upper signs hold for $i=1,3,$ and compute the 
corresponding $X_i=\ket{\phi(\alpha)}\bra{\psi_i},$
which leads to the four nonlinear correction terms 
$\NL_i := \mean{X_i^{T_B}}\mean{(X_i^{T_B})^\dagger}$
with
\begin{align}
\NL_{1/2} &= 
\frac{1}{32}
\big[
\big(
(c \pm s)(1 \pm x_1 x_2 \pm y_1 y_2 + z_1 z_2)
\nonumber
\\
+& (c \mp s)(z_1 + z_2)
\big)^2
+ 
\big(
(c \mp s)(x_1y_2-y_1x_2)
\big)^2
\big],
\nonumber
\\
\NL_{3/4} &= 
\frac{1}{32}
\big[
\big(
(c \pm s)(x_1 \pm x_2)
+
(c \mp s)(z_1 x_2 \pm x_1 z_2)
\big)^2
\nonumber
\\
+&
\big(
(c \mp s)(y_1 \mp y_2)
+
(c\pm s)(y_1 z_2 \mp z_1 y_2)
\big)^2
\big].
\label{nl10}
\end{align}
With these nonlinearities, the nonlinear witnesses 
$\FF^{(1)},\FF^{(2)},$ and $ \FF^{(3)}$ according 
to Eqs.~(\ref{nl3}, \ref{nl4}, \ref{nl7})
can directly be written down. Note that, especially for 
$c=\pm s,$ several quadratic terms in the $\NL_i$ 
can be measured with the same measurements as in 
Eq.~(\ref{nl9}). Thus their implementation requires 
no extra effort.

\begin{figure}[t]
\centerline{\epsfxsize=0.9\columnwidth
\epsffile{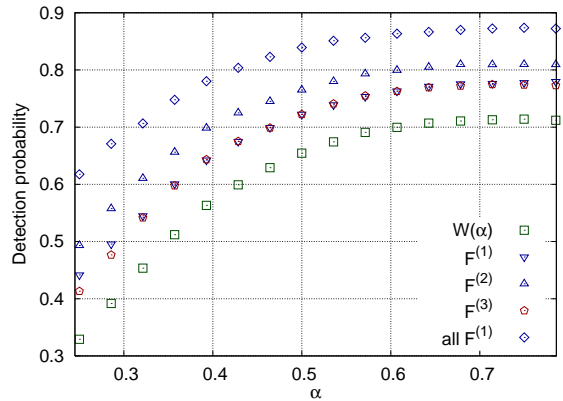} }
\caption{Probability of detecting a state with a witness and with 
nonlinear criteria. See text for details.}
\end{figure}

To investigate the power of these new criteria, 
we consider  the family of states
$\vr(\alpha)= (\eins -\ketbra{\phi(\alpha)}^{T_B})/3,$
which are separable, but they lie on the boundary of the 
set of separable states, since 
$Tr(\WW(\alpha)\vr(\alpha))=0.$ 
For different $\alpha$ we generate randomly states 
$\vr_r$ with $\Vert \vr(\alpha) - \vr_r \Vert \leq 0.2$
in Hilbert Schmidt norm \cite{karolz}. For the entangled 
states in this ball we determine the probability of 
detecting it via the witness $\WW(\alpha)$ and via a 
nonlinear criterion. First, we consider $\FF^{(1)}$ 
as in Eq.~(\ref{nl3}), 
using $\ket{\psi_2}$ and $2 \cdot \NL_2.$ Then, we consider $\FF^{(2)}$ as in 
Eq.~(\ref{nl4}), using all nonlinearities. We also test 
the criterion in Eq.~(\ref{nl5}), corresponding to 
all possible $\FF^{(1)}.$ Finally, we investigate 
$\FF^{(3)}$ from Eq.~(\ref{nl7}) using  
$\ket{\psi_3}, \ket{\psi_4}$ and $s_\phi=c^2.$   
The results are shown in Fig.~1. Clearly, the nonlinear 
criteria improve the witness significantly \cite{lurremark}.

Let us now discuss generalizations of our approach to other 
entanglement witnesses, which are not related to the PPT 
criterion. This can be done via {\it positive maps} 
\cite{hororeview}. Let $\HH_B$ and $\HH_C$ be Hilbert 
spaces and let $\BB(\HH_i)$ denote the linear operators 
on it. A linear map $\Lambda: \BB(\HH_B) \rightarrow \BB(\HH_C)$ 
fulfilling $\Lambda(X) \geq 0$ for $X \geq 0$ and 
$\Lambda(X^\dagger)=\Lambda(X)^\dagger$ is called {\it positive} (P).
A positive map is {\it completely positive} (CP) when
for an arbitrary $\HH_A$ the map $\II_A \otimes \Lambda$ is P,
otherwise, it is positive, but not completely positive (PnCP).
Here, $\II_A$ denotes the identity on $\BB(\HH_A).$ 
For example, the transposition is PnCP: While $X \geq 0$ implies 
$X^T \geq 0$ the {partial} transposition does not preserve the 
positivity of a state. 

Indeed, it has been shown \cite{ppthoro, hororeview} that a state 
$\vr \in \BB(\HH_A) \otimes \BB(\HH_B)$ is separable iff 
for all P maps $\Lambda$ the relation
$ \II_A \otimes \Lambda (\vr) \geq 0.$
holds. Consequently, if $\vr$ is entangled there must 
be a trace decreasing PnCP map $\Lambda$ where $\II_A \otimes \Lambda (\vr)$ 
has a negative eigenvalue $\lambda_-$ and a corresponding 
eigenvector $\ket{\phi}.$ Taking $(\II_A \otimes \Lambda)^+$ 
as the adjoint of the map $(\II_A \otimes \Lambda)$ with respect 
to the scalar product $Tr(X^\dagger Y)$ a witness
detecting $\vr$ is given by
\be
\WW= (\II_A \otimes \Lambda)^+ (\ketbra{\phi}),
\label{nl14}
\ee
since we have 
$Tr[\rho \WW] = Tr[\rho (\II_A \otimes \Lambda)^+(\ketbra{\phi})]
= Tr[\II_A \otimes \Lambda (\vr) \ketbra{\phi}] = \lambda_-.$ 
Replacing the partial transposition by the map 
$(\II_A \otimes \Lambda)^+$ this witness can then 
be improved as in Observation 1.

For an arbitrary witness, we make use of the Jamio{\l}kowski 
isomorphism \cite{jamiol, hororeview} between operators
and maps. According to this, an operator $E$ on 
$\BB(\HH_B) \otimes \BB(\HH_C)$ corresponds to a map
$ \varepsilon : \BB(\HH_B) \rightarrow \BB(\HH_C)$
acting as  $\varepsilon (\vr) = Tr_B (E \vr^T \otimes \eins_C).$
Conversely, we have
$
E = (\II_{B'} \otimes \varepsilon)(\ketbra{\phi^+})
$
where $\HH_{B'}\cong \HH_{B}$ and $\ket{\phi^+}= \sum_i \ket{ii}$
is a maximally entangled state on  $\HH_{B'} \otimes \HH_B.$
The key point is that if $E$ is an entanglement witness, then 
$\varepsilon$ is a PnCP map \cite{hororeview, tracepreserving}. 
Hence, any witness can be written as in Eq.~(\ref{nl14}) and 
finally we have:

{\bf Theorem 1.} Any bipartite entanglement witness can be improved 
by nonlinear corrections using the methods of Observation 1.

In conclusion, we have shown that all bipartite entanglement 
witnesses can be improved by nonlinear expressions. 
These nonlinear witnesses are straightforward to calculate and 
can also be directly implemented in experiments, often without 
any extra effort. It is tempting to extend these constructions 
to the multipartite scenario.
Here, this challenge remains undone. 

We thank H.J. Briegel, M. Curty, J. Eisert, P. Hyllus, B. Kraus, 
M. Lewenstein, M. Piani and G. T\'oth for discussions. This work 
has been supported by the FWF, the DFG (Emmy-Noether-Programm) and the EU 
(OLAQI, PROSECCO, QAP, QUPRODIS, SCALA).

{\it Appendix ---} Here we prove Observation 2. 
(a) First, note that 
$Y = \vr^{T_B}\ketbra{\phi}\vr^{T_B} =: \ketbra{\eta}$
is projector onto a not normalized state $\ket{\eta}.$ 
To derive a criterion for the strength of $\FF^{(1)}$ 
we have to maximize 
$
\mean{X^{T_B}} \mean{(X^{T_B})^\dagger}/s(\psi)
= \braket{\psi}{\eta}\braket{\eta}{\psi}/s(\psi)
$ 
for a given $\ket{\eta}$ over all $\ket{\psi}.$
Let $\ket{\eta}=\sum_i a_i \ket{i i}$ and
$\ket{\psi}= \sum_i b_i \ket{\tilde{i} \tilde{i}}$
be the Schmidt decompositions of $\ket{\eta}$ and 
$\ket{\psi},$ with decreasingly ordered Schmidt
coefficients. We first show that for fixed Schmidt 
coefficients $b_i$ it is optimal to take 
$\ket{\tilde{i} \tilde{i}} = \ket{i i}.$ 
We have
$
|\braket{\eta}{\psi}|
= 
|\sum_{ij} a_i b_j \braket{ii}{\tilde{j}\tilde{j}}|
=
|\sum_{ij} a_i b_j U^A_{ij} U^B_{ij}|
$
where  $U^A_{ij}= \braket{i}{\tilde{j}}^A$ and 
$U^B_{ij}= \braket{i}{\tilde{j}}^B$ are unitary.
Defining $\fa_{ij}= \sqrt{a_i}\sqrt{b_j}U^A_{ij}$
and $\fb_{ij}= \sqrt{a_i}\sqrt{b_j}U^B_{ij}$ this 
can be written as a scalar product which is maximal, when
$\fa_{ij}, \fb_{ij}$ are parallel. Due to the 
Cauchy-Schwarz inequality, we can then assume 
without loosing generality
$ |\braket{\eta}{\psi}| \leq \sum_{ij} a_i b_j |U^A_{ij}|^2.$
Since $U^A_{ij}$ is unitary, $|U^A_{ij}|^2$ is doubly 
stochastic, i.e. its row and column sums equal one. Due to 
Birkhoff's theorem \cite{hornjohnson} it can be written as 
a convex combination of permutation matrices.
For a permutation $\pi$ we have
$\sum_{i} a_i b_{\pi(i)} \leq \sum_{i} a_i b_i$ due to the 
ordering of the $a_i$ and $b_j$. This implies finally
$|\braket{\eta}{\psi}| \leq \sum_{i} a_i b_i$ where equality
is achieved when $\ket{\tilde{i} \tilde{i}} = \ket{i i}.$
Having fixed $\ket{\tilde{i} \tilde{i}} = \ket{i i}$ we only 
have maximize $\sum_i (b_i/b_0) a_i$ over all $b_i.$ This is 
maximal when $(b_i/b_0)=1$ for all $i,$ which corresponds to
a maximally entangled  $\ket{\psi}= 1/\sqrt{d} \sum_i \ket{i i}.$
But then
$\braket{\psi}{\eta}\braket{\eta}{\psi}/s(\psi) = (\sum_i a_i)^2$
corresponds to the right hand side of Eq.~(\ref{nl5}).
(b) First, note that for an arbitrary basis $\ket{\psi_i}$ in  
Eq.~(\ref{nl4})
$\sum_{i=1}^K \mean{X_i^{T_B}} \mean{(X_i^{T_B})^\dagger} = 
Tr(Y) = \bra{\phi}(\vr^{T_B})^2 \ket{\phi}.$ This shows  
Eq.~(\ref{nl6}). 
(c)  Assume that $\vr$  fulfills Eq.~(\ref{nl6}). Since $Y$ 
is of rank one, there is a single $\ket{\psi'}$ such that 
$Tr(Y)=\bra{\psi'}Y\ket{\psi'}.$ So, if we take 
$X=\ket{\phi}\bra{\psi'}$ then the witness $
\FF = \mean{\ketbra{\phi}^{T_B}} - \mean{X^{T_B}} 
\mean{(X^{T_B})^\dagger}
$
detects $\vr$. The witness
$
\FF'=
\mean{\ketbra{\phi}^{T_B}} - 
\mean{X^{T_B}} \mean{(X^{T_B})^\dagger} /{s(\psi')}
$
is stronger, and detects it as well. Finally, we have to show 
that Eq.~(\ref{nl5}) is never valid for a PPT state $\vr.$
In view of the proof of (a), it suffices to show
$Q:=\bra{\phi} \vr^{T_B} \ket{\phi} 
- \bra{\psi} \vr^{T_B} \ket{\phi} \bra{\phi} \vr^{T_B} \ket{\psi}
/{s(\psi)} \geq 0$ for arbitrary $\ket{\phi}$ and maximally 
entangled $\ket{\psi}.$ Since $\vr^{T_B} \geq 0$ we can 
define $R=\sqrt{\vr^{T_B}}\ketbra{\phi}\sqrt{\vr^{T_B}}$
and $S=\sqrt{\vr^{T_B}}\ketbra{\psi}\sqrt{\vr^{T_B}}/{s(\psi)},$
then $Q=Tr(R)-Tr(RS)= Tr(R(\eins-S)).$ Since $R \geq 0$ and 
$S \geq 0$ it suffices to show that $Tr(S)<1$, then 
$(\eins-S) \geq 0$ follows and finally $Q \geq 0.$ We have 
$Tr(S) = \bra{\psi}\vr^{T_B}\ket{\psi}/{s(\psi)}.$ Now, 
we use the known fact that a witness like 
$\WW=s(\psi)\eins - \ketbra{\psi}$ where $\ket{\psi}$
is maximally entangled, detects no PPT states 
\cite{sanpera}. Since $\vr^{T_B}$ is PPT, this 
implies that $Tr(S)<1.$ 
$\qed$


\begin{thebibliography}{99}

\bibitem{witnessexp} 
For experiments using witnesses see 
\cite{mohamed} and 
M.~Barbieri {\it et al.}, 
Phys. Rev. Lett. {\bf 91}, 227901 (2003);
K.J. Resch, 
P. Walther and A.Zeilinger,
{\it ibid.} {\bf 94}, 070402 (2005); 
J. Altepeter {\it et al.},
{\it ibid.}  {\bf 95}, 033601 (2005);
H.~Mikami {\it et al.},
{\it ibid.} {\bf 95}, 150404 (2005); 
N. Kiesel {\it et al.},
{\it ibid.} {\bf 95}, 210502 (2005); 
D.~Leibfried {\it et al.},
Nature (London) {\bf 438}, 639 (2005);
H.~Häffner {\it et al.},
{\it ibid.}  {\bf 438}, 643 (2005).


\bibitem{mohamed}
M.~Bourennane {\it et al.},~Phys.~Rev.~Lett.~{\bf 92},~087902~(2004).

\bibitem{witnessth} For theoretical studies see 
\cite{ppthoro, optimization, sanpera, pmap} and 
B. Terhal, 
Phys. Lett. A {\bf 271}, 319 (2000);
D. Bru\ss~{\it et al.}, 
J. Mod. Opt. {\bf 49}, 1399 (2002); 
G. T\'oth and O. Gühne, 
Phys. Rev. Lett. {\bf 94}, 060501 (2005);
F. Brand\~ao, 
Phys. Rev. A {\bf 72}, 022310 (2005).

\bibitem{ppthoro}
M. Horodecki, P. Horodecki, and R. Horodecki,  
Phys. Lett. A {\bf 223}, 1 (1996).

\bibitem{optimization}
M. Lewenstein {\it et al.}, 
Phys. Rev. A {\bf 62}, 052310 (2000).

\bibitem{sanpera}
A. Sanpera, D. Bru{\ss}, and M. Lewenstein,
Phys. Rev. A {\bf 63}, 050301 (2001).

\bibitem{pmap}
M. Lewenstein {\it et al.}, 
Phys. Rev. A {\bf 63}, 044304 (2001).

\bibitem{nonlinear}
L. Duan {\it et al.}, 
Phys. Rev. Lett. {\bf 84}, 2722 (2000);
N.~Korolkova {\it et al.}, 
Phys. Rev. A {\bf 65}, 052306 (2002);
P.~van Loock and A. Furusawa, 
{\it ibid.} {\bf 67}, 052315 (2003);
G.~T\'oth, C. Simon, and  J.I. Cirac,
{\it ibid.} {\bf 68}, 062310 (2003);
S. Yu {\it et al.}, 
Phys. Rev. Lett. {\bf 91}, 217903 (2003); 
E. Shchukin and W. Vogel
{\it ibid.} {\bf 95}, 230502 (2005);
F.A. Bovino {\it et al.}, 
{\it ibid.} {\bf 95}, 240407 (2005);
A.~Serafini, 
{\it ibid.} {\bf 96}, 110402 (2006).


\bibitem{uffink}
J. Uffink, 
Phys. Rev. Lett. {\bf 88}, 230406 (2002).
 
\bibitem{lurs}
H.F.  Hofmann and S. Takeuchi, 
Phys. Rev. A {\bf 68}, 032103 (2003).


\bibitem{aip} O.~Gühne and M. Lewenstein,
AIP Conf. Proc. {\bf 734}, 230 (2004);
G. T\'oth and O. Gühne,
Phys. Rev. A {\bf 72}, 022340~(2005).

\bibitem{hyllus} 
V. Giovannetti {\it et al.}, 
Phys. Rev. A {\bf 67}, 022320 (2003);
P. Hyllus and J. Eisert,
New J. Phys. {\bf 8}, 51 (2006).

\bibitem{uncertaintyprl}
O. Gühne,
Phys. Rev. Lett. {\bf 92}, 117903 (2004).

\bibitem{ppt}
A. Peres,  
Phys. Rev. Lett. {\bf 77}, 1413 (1996). 

\bibitem{jarek} Note the similarity of this problem
to  J.K.~Korbicz {\it et al.}, Phys. Rev. Lett. 
{\bf 94}, 153601 (2005). 


\bibitem{karolz}
K. \.Zyczkowski and H.-J.~Sommers, J. Phys. A {\bf 34}, 7111 (2001).


\bibitem{lurremark} 
Note that from the local uncertainty relations \cite{lurs} 
it follows that
$ 
\FF=\mean{\WW(\pi/4)} 
- \big[(x_1+x_2)^2 + (y_1+y_2)^2+ (z_1+z_2)^2 \big]/8
$
is also an improvement of $\WW(\pi/4)$ \cite{aip}. 
Using Eq.~(\ref{nl4}) one gets a similar expression 
$\FF^{(2)},$ however, the numerics shows
that $\FF^{(2)}$  is slightly better than $\FF.$

\bibitem{hororeview}
For a review see
M. Lewenstein, {\it Quantum information theory}, available at
{http://www.itp.uni-hannover.de/tqowww/download.php};
M. Horodecki, P.~Horodecki, and R.~Horodecki, quant-ph/0109124.

\bibitem{jamiol}
A. Jamio{\l}kowski, Rep. Mat. Phys. {\bf 3}, 275 (1972).

\bibitem{tracepreserving}
We assume without loosing generality 
that $\varepsilon^+$ is trace decreasing, 
we can always rescale $E$ to obtain this.

\bibitem{hornjohnson} See Theorem 8.7.1 in
R.A. Horn and C.R. Johnson, {\it Matrix analysis}
(Cambridge University Press, 1999).

\end{thebibliography}
\end{document}